\documentclass{iaus}
\usepackage{graphicx}

\title[Masers in star forming regions] %% give here short title %%
{Masers in star forming regions}

\author[Anna Bartkiewicz \& Huib Jan van Langevelde]   %% give here short author list %%
{Anna Bartkiewicz$^1$
 \and Huib Jan van Langevelde$^{2,3}$}

\affiliation{$^1$Toru\'n Centre for Astronomy, Nicolaus Copernicus
University,\\ Gagarina 11, 87-100 Toru\'n, Poland, email {\tt annan@astro.uni.torun.pl} \\[\affilskip]

$^2$Joint Institute for VLBI in Europe,\\
Postbus 2, 7990 AA Dwingeloo, The Netherlands, email {\tt langevelde@jive.nl} \\[\affilskip]

$^3$Sterrewacht Leiden, Leiden University, \\
Postbus 9513, 2300 RA Leiden, The Netherlands}

\pubyear{2012}
\volume{287}  %% insert here IAU Symposium No.
\pagerange{1--10}
\jname{Cosmic Masers -- from OH to H$_{\rm o}$}
\editors{R.S. Booth, E.M.L. Humphreys \& W.H.T. Vlemmings, eds.}
\begin{document}

\maketitle

\begin{abstract}
Maser emission plays an important role as a tool in star formation studies.
It is widely used for deriving kinematics, as well as the physical conditions of
different structures, hidden in the dense environment very close to the young
stars, for example associated with the onset of jets and outflows. We will summarize here
the recent observational and theoretical progress on this topic since
the last maser symposium: the IAU Symposium 242 in Alice Springs.
\end{abstract}

\firstsection % if your document starts with a section,
              % remove some space above using this command.
\section{Introduction}

Cosmic masers are known as a unique tool in star-formation studies and are one of the first
observed signpost of high-mass star formation, particularly the hydroxyl (OH), water
(H$_2$O) and methanol (CH$_3$OH) masers, that are common and intense. This
is demonstrated by the many results presented in this volume. 

At the last IAU Symposium 242 in Alice Springs, \cite[Fish (2007)]{Fish07} 
summarized the relation between masers and star-formation in the following way {\it "maser observations are at
the vanguard of star formation research: yesterdays observations can be
explained by complementary data and theory today, and todays observations lay the
groundwork for the breakthroughs that will be achieved in the context of
tomorrow."} In this
review we will summarize some of the achievements and discoveries in the
area of star-formation masers presented in the literature since  
the Australian Symposium. It is important to evaluate what 
"yesterday's tomorrow" has unveiled in the area of star formation and 
identify the possible "today's tomorrow" breakthroughs. 

\section{Population studies}

It was relatively well established from earlier surveys of masers in our Galaxy 
that massive star-forming regions can be associated with OH, H$_2$O and
Class~II CH$_3$OH masers
(e.g., \cite[Caswell et al. 1995]{caswell95}, \cite[Szymczak et al.
2002]{szymczak02}). However, there is still a
great need for verifying what the relation is between specific stages and classes of star formation and
different masers. For this, complete and ever more sensitive surveys with better astrometric
precision are most valuable. For example, since our previous meeting \cite[Green et al. (2009)]{green09}
discovered that high-mass star
formation (HMSF) is present in both the far and near 3~kpc arms through 49 detections of 6.7~GHz
methanol masers. The Red MSX Sources (RMS) based survey by \cite[Urquhart et al.
(2009)]{urquhart09} investigated the
statistical correlation of water masers with early-stage of massive
star-formation. They found similar detection rates for UC~H~{\small II}
regions and MYSOs, suggesting that the conditions needed for maser activity are
equally likely in these two stages of star formation. 

Nowadays, more instruments have become available, especially to focus systematically on different maser
transitions like the higher excited methanol masers from both Class I (collisional excitation) and
II (radiative excitation) methanol, as well as silicon monoxide (SiO), ammonia (NH$_3$) and
formaldehyde (H$_2$CO) masers. Details are presented by e.g.\ Kurtz, Kalenskii, 
Voronkov, Sjouwerman, Wootten, Booth, Kim, Pestalozzi, Brogan and their collaborators in these proceedings.

In addition, yesterday's key-questions, which were deemed essential in order to use masers for
studying the physics of star formation, are still not fully answered:
\begin{itemize}
\item{Is there an evolutionary sequence based on maser occurrence?}
\item{Are Class I and Class II methanol masers associated?}
\item{Where, when and how exactly do masers arise?}
\item{What physical conditions are needed to produce the maser(s)?}
\end{itemize}

These questions require systematic studies of a large number of sources, possibly at high angular
resolution, as well as observations of specific
sources using multi-wavelength observations, in order to converge 
and refine our hypotheses. 
For example, single-dish studies suggested at first that the methanol Class~I and II
are coincident, but later interferometric images showed they are not co-spatial on arcsecond 
scales, even though they may be driven by the same YSO \cite[(Cyganowski et al.
2009)]{cyganowski09}. 

\subsection{Methanol masers, the most widespread masers in HMSFRs}

Methanol masers have been widely studied, particularly after the discovery of the wide\-spread, bright
6.7 GHz transition \cite[(Menten 1991)]{menten91}. 
Many transitions have been found to maser from both the A and E types and they have been classified into 
Class~I and Class~II, which are collisionally and radiatively excited, 
respectively \cite[(Cragg et al. 1992)]{cragg92}. In general, the 
Class~I (e.g., 36 and 44 GHz lines) are likely associated with outflows
(lying further from the central objects) while the Class~II (e.g., 6.7 and
12.2~GHz transitions) often coincide with hot
molecular cores, UC H~{\small II} regions, OH masers and near-IR sources.

The most common Class~II masers are 6.7 and 12.2~GHz lines that 
according to the models (e.g., 
\cite[Cragg et al. 2005]{cragg05}) should co-propagate quite often, as confirmed by 
observations (e.g., \cite[Breen et al. 2011]{breen11}). Both masers
are strongly associated with HMSFRs and enable us to probe the dense environments
where stars are being born. A large sample of 113 sources with known 6.7~GHz
masers and 1.2-mm dust clumps was searched at 12.2~GHz by \cite[Breen et al.
(2010b)]{breen10b}.
These authors found out that when the 6.7~GHz emission is more luminous, the
the evolutionary stage of the central object tends to be more advanced, while 
also the 12.2~GHz is often associated with more evolved regions. Based on this 
an evolutionary sequence for masers associated with massive
star formation regions was proposed, which is consistent with conclusions of e.g.\
the survey of Class I by \cite[Pratap et al. (2008)]{pratap08}. However, some discussion
continues to refine this relation, e.g.\ \cite[Fontani et al. (2010)]{fontani10} and \cite[Chen et al.
(2011)]{chen11}.

\subsubsection{Class~I and Class~II methanol masers}

One of the key research areas over the past few years has been on relatively rare methanol masers.
\cite[Voronkov et al. (2010b)]{voronkov10b} found two new detections of
9.9~GHz Class I masers. To date we know of 5 masers at 9.9~GHz 
(additionally from \cite[Slysh et al. 1993]{slysh93}, Voronkov et al. 2006, 2011). The detection rate is 
likely so low because of the strong dependence of the maser brightness on the 
physical conditions (\cite[Sobolev et al. 2005]{sobolev05}). This maser is believed to trace shocks
caused by different phenomena (e.g., expanding H~{\small II} regions, outflows).
A particularly interesting case is
G331.13$-$00.24, which shows periodic variability at 6.7~GHz with a period of
500~days (\cite[Goedhart et al. 2004]{goedhart04}). There is an obvious urgency to verify whether 
the variations of both
lines correlate, pointing to a common origin of the seed radiation and providing
an estimate of the physical conditions for that. For more details see Voronkov et al.\ (these proceedings).

There are now also first arcsecond-resolution images of the 36~GHz methanol masers in
HMSFRs thanks to the upgrades of both ATCA and EVLA, e.g.\ in M8E 
(\cite[Sarma \& Momjian 2009]{sarma09}), Sgr~A (\cite[Sjouwerman et al. 2010b]{sjouwerman10b}),
G309.38$-$0.13 (\cite[Voronkov et al. 2010a]{voronkov10a}) and DR21 \cite[(Fish et al.
2011b]{fish11b}). In the latter case it was found 
that surprisingly the Class~I 36~GHz and 229~GHz masers appear in close proximity (also
in velocity) with the Class~II 6.7~GHz maser, while the 44~GHz Class~I masers is absent. 
According to the model by \cite[Voronkov et al. (2005)]{voronkov05} 
such cases require an intermixed environment of dust and gas at a lowish 
temperature of $\approx$60~K.

One may wonder whether we have come closer to answering the question {\it "when do Class~I masers appear?"}.
\cite[Chen et al. (2011)]{chen11} searched 192 EGOs (the candidates
associated with ongoing outflows) for 95~GHz methanol 
masers, resulting in a 55~per~cent detection rate. These detections are likely associated with the 
redder GLIMPSE point-source colors. There are two possible 
explanations, either the Class~I objects are associated with lower stellar masses or 
they are associated with more than one evolutionary phase during high-mass star
formation, apparently contradicting the most straightforward schemes \cite[(e.g.\ Breen et al., 2010b)]{breen10b}. \cite[Marseille et al. (2010)]{marseille10} compared the physical conditions 
by observing several molecular tracers in both weak and bright mid-IR emitting massive 
dense cores. The methanol Class~I maser at
84.5~GHz was found to be strongly anti-correlated with the 12~$\mu$m source brightness,
leading to an interpretation that these represent more embedded mid-IR sources with a spherically
symmetric distribution of the envelope material. 

\cite[Ellingsen et al. (2011)]{ellingsen11} searched for rare and weak
methanol masers at 37.7, 38.3, 38.5~GHz Class~II methanol masers towards 70 HMSFRs.
They detected 13 at 37.7~GHz and 3 at 38.3/5~GHz and found that the 37.7~GHz
masers are associated with the most luminous 6.7 and 12.2~GHz masers,  
likely representing a short (of 1000--4000 years) period in an advanced stage of the evolution.
Therefore, the 37.7~GHz methanol masers may be called the
{\it horsemen of the apocalypse} for the Class~II methanol maser phase.

\subsubsection{The morphology of 6.7~GHz masers}

More sensitive VLBI surveys have led to the discovery of more complex 6.7~GHz maser structures, 
including several that show a ring-like morphology (Bartkiewicz et al. 2005, 2009). 
Kinematics of the maser spots revealed that outflow/infall dominates over
the possible Keplerian rotation in a disc. A similar morphology with a similar
kinematic signature was found in the well-known HMSFR Cep~A, 
where, due to additional constraints on the orientation, 
the radial motions are more likely resulting from infall (\cite[Torstensson et al.
2011a]{torstensson11a}, \cite[Sugiyama et al. 2008]{sugiyama08}). Moreover, 
it seems that the magnetic field plays a role in
shaping this morphology (\cite[Vlemmings et al. 2010]{vlemmings10}). 
Such ring-like  
characteristics were also seen in water masers associated with slightly more
advanced stages, where masers were likely tracing an accretion disc or its
remnant \cite[(Motogi et al. 2011a)]{motogi11a}. 
\cite[Torstensson et al. (2011b)]{torstensson11b} 
analysed some of these ring-like maser sources using thermal emission at
arcsec scale and found that mostly the distribution of the methanol gas peaks at the 
maser position with the larger scale gas showing a modest outflow velocity. 
They argued that the methanol gas has a single origin in these sources, possibly associated 
with an accretion shock. ALMA resolution is necessary 
for probing the regions of interest at size scales of 1000~AU.

For all of these studies it is important to remember that the VLBI techniques resolves out some
of the emission.  
\cite[Pandian et al. (2011)]{pandian11} noted that more complex morphologies 
and often larger structures become apparent when using shorter baseline
interferometers (EVLA, MERLIN) compared to VLBI, analyzing a study of 72 sources from the Arecibo Methanol
Masers Galactic Survey. Thus, the 29\% detection of the methanol rings
of \cite[Bartkiewicz et al. (2009)]{bartkiewicz09} may be biased by
observational effects. Similarly, \cite[Cyganowski et al. (2009)]{cyganowski09} also 
noted that shorter baselines observations resulted in more complex and extended emission
for two targets from the EVN sample of 
\cite[Bartkiewicz et al. (2009)]{bartkiewicz09}. 
On the other hand, comparing \cite[Pandian et al. (2011)]{pandian11} and
\cite[Bartkiewicz et al. (2009)]{bartkiewicz09} results, we note that 
in three out five cases the emission is very similar on EVN and MERLIN images, but
this does not include any of the ring sources.

\section{Gas kinematics through proper motion studies of masers}

There has also been significant progress on maser proper motion
studies at mas scale. These result from multi-epoch observations that often have 
two simultaneous objectives: proper motions of the maser features in order to derive the 
kinematics of the gas in the direct environment of the YSOs and accurate direct distances 
by means of detecting the parallax. The
parallax measurements are summarized and presented in this volume by e.g., Reid, 
Honma et al., Nagayama et al., Choi et al.\ (these proceedings)

Here we focus on the dynamics studies. In a series of papers, Moscadelli et al. (e.g., 2007, 2011a) 
demonstrated the power of multi-epoch VLBI for tracing the 3D kinematics close to an YSO 
towards nine well-studied HMSFRs. Combining observations of
22~GHz water and 6.7~GHz methanol masers within a time-span of a few years
they detected various motions such as outflow, rotation, infall, all happening in the direct
environments of these YSOs. They pointed out that these velocity gradients on mas scale
still reflect large-scale (100-1000~AU) motions (\cite[Moscadelli et al.
(2011b)]{moscadelli11b}). 
In some cases such studies can constrain the YSO position and mass.
For example, \cite[Goddi et al. (2011)]{goddi11} directly measured these different phenomena
going on within 400~AU from the high-mass protostar AFGL~5142; the gas infall 
is traced by the 3D velocities of the methanol masers, while
a slow, massive, collimated, bipolar outflow is detectable through the water
masers. Very detailed dynamics were registered by 
\cite[Torrelles et al. (2011)]{torelles11} who used multi-epoch data of water
masers towards Cep~A HW2, noticing morphological changes at scales of 70~AU in a time-span of
5~years. They also argued that the R5 expanding bubble structure has been dissipating in the 
circumstellar medium and that a slow, wide-angle outflow at the scale 
of 1000~AU co-exists with the well-known high-velocity jets. 

In addition to these methanol and water maser observations, there
are unique data from SiO masers, although they are quite rare around YSOs. \cite[Matthews
et al. (2010)]{matthews10} observed the Orion Source~I at both 43.1 and 42.8~GHz transitions,
resulting in a most detailed view of the inner 20-100~AU of a MYSO. 
The SiO masers lie in an X-shaped structure, with clearly separated blue- and
redshifted emission, while bridges of intermediate-velocity emission connects both
sides. They proposed that these masers are related to a wide-angle bipolar wind 
emanating from a rotating, edge-on disc. This is providing direct evidence of 
the formation of a MYSO via disc-mediated accretion.
Other examples and more explanations can be found in contributions 
by e.g.\ Sanna et al., Goddi et al., Sawada-Satoh et al., Sugiyama et al.\ (these proceedings).

\section{Physical conditions for maser emission}

In order to answer the key question {\it where, when and how exactly do
masers arise?}, we must probe the physical conditions in which they form. Such studies 
concern multiple maser transitions and studying the masing regions at a wide 
range of wavelengths. Both surveys of a large number of sources, as well as
detailed individual source observations are needed to complete the scenario of maser
formation. A very good example is the result obtained by \cite[Cyganowski et al. (2008)]{cyganowski08}
that Class~I and II methanol masers coincide with so-called 
extended green objects (EGOs) which are indicators of outflows and 
are a promising starting point for identifying MYSOs (Cyganowski,
these proceedings). 

\cite[Breen et al. (2010a)]{breen10a} investigated the
OH/H$_2$O/CH$_3$OH relation for a large sample of HMSFRs and noticed 
a closer similarity of the velocities of OH and methanol masers than 
of either of these species compared to the water maser peak
velocity. In spite of the different pumping schemes of water and methanol
masers, they both show a similar, 80\% detection rate association with OH sources.  
It also has been found by comparing high-luminosity masers with low-luminosity ones 
that the high brightness ones are related to lower NH3(1,1) excitation temperatures, 
smaller densities, but three times larger column densities. Moreover, the
high-luminosity sources are associated with 10 times more massive molecular cores, larger outflows
and their internal motions are more pronounced \cite[(Wu et al. 2010)]{wu10}. 
Interestingly, \cite[Pandian et al. (2010)]{pandian10} showed that the continuum of the counterparts 
of 6.7~GHz methanol masers is consistent with rapidly accreting massive YSOs
($>$0.001~M$_\odot$~yr$^{-1}$) by constraining their SEDs. Only a minority of the
sample, 30\%, coincided with H~{\small II} regions
that are usually ultra- or hyper compact. The latter 
was also confirmed by \cite[S\'anchez-Monge et al. 
(2011)]{sanchez11} and \cite[Sewilo et al. (2011)]{sewilo11}. 
Indeed the majority of 6.7~GHz masers seems to appear before the H~{\small II} 
stage of MYSOs, as was suggested by earlier studies, e.g., \cite[Walsh et al.
1998]{walsh98}. Alternatively, we may still not have been able to reach the proper
sensitivity for such conclusions.

Studies of specific sources in multiple maser transitions and
their counterparts in other tracers are of special value. In the well known ON~1 source 
OH transitions at 1.612, 1.665, 1.667, 1.720, 6.031 and 6.035~GHz lie in a
similar region as 6.7 GHz methanol masers. \cite[Green et al.
(2007)]{green07} concluded that they possibly trace a shock front in the 
form of a torus/ring around the YSO. That interpretation is also supported by polarization 
angles and velocity gradients. In the HMSFR NGC~7538 IRS~1 
new masers at 12.2~GHz were found (\cite[Moscadelli et al.
2009]{moscadelli09}) and in addition, 23.1~GHz Class~II methanol masers
were accurately registered (\cite[Galv\'an-Madrid et al. 2010]{galvan10}). They appear closely
associated with 4.8~GHz H$_2$CO masers, indicating that the conditions must be similar
for both of these relatively rare masers. It is possible that they are excited by the free-free 
emission from an H~{\small II} region. However, surprisingly, they are not accompanied by any 6.7
or 12.2~GHz methanol masers.

Although we are collecting more and more information, the origin of maser structures
in high-mass star formation is still not clear. A long-term question
{\it what do linearly distributed methanol masers trace, an edge-on disc or an
outflow?} is still open. \cite[De Buizer et al. (2009)]{debuizer09} showed 
that orientations of SiO outflows were not consistent with the
methanol masers delineating a disk orientation. Moreover, for the methanol rings
the proposed morphology could generally not be confirmed by infrared high resolution
imaging (De Buizer et al., these proceedings). \cite[Beuther et al.
(2009)]{beuther09}, using NH$_3$ as a tracer towards methanol Class~II masers 
found that if Keplerian accretion disks exist, they should be confined to regions 
smaller than 1000~AU. Therefore, 
ALMA-resolution observations are really needed in order to reach the relevant scales in 
the direct environment of MYSOs. 

\section{Masers as a signpost of star formation}

Masers are readily usable as a diagnostic in complex SFRs, for example as indicators that
star formation has begun. \cite[Purcell et al. (2009)]{purcell09} investigated
the NGC~3576 region and verified the evolutionary status of the various molecular
components. Water masers were found towards the NH$_3$ emission peaks, lying
in the arms of the filament. In the HMSFR G19.61$-$0.23 water masers trace
the outflow/jet associated with the most massive core, SMA~1, also traced by
H$^{13}$CO$^{+}$ emission (\cite[Furuya et al. 2011]{furuya11}). The
massive cold dense core G333.125--0.562 showed water 
and methanol masers, as well as SiO thermal emission, but 
remained undetected at wavelengths shorter than
70~$\mu$m \cite[(Lo et al. 2007)]{lo07}. Moreover, 44~GHz methanol masers coincide
with presumably masing 23~GHz NH$_3$ emission in the EGO
G35.03$+$0.35 \cite[(Brogan et al. 2011)]{brogan11}. The latter project is  
based on simultaneous observations of continuum emission and a comprehensive set of lines,
something that has become possible with the Expanded Very Large Array (EVLA) and
should contribute significantly to our understanding of star formation (Brogan et al., this volume). 

\section{Variability of masers}

The 6.7 and 12.2~GHz methanol masers have been monitored and unexpectedly
periodic variations were discovered from some masers in HMSFRs (e.g.\ 
\cite[Goedhart et al. 2003]{goedhart03}). Such studies are 
possible with single-dish observations and often require 
long-term commitments. Monitoring can provide important clues about which phenomena 
are responsible for the variability, but also about the more general physical conditions in the masing
regions or the background radiation field. 

Recently, \cite[Goedhart et al. (2009)]{goedhart09} summarized nine years 
of monitoring of G12.89$+$0.49 at the 6.7 and 12.2~GHz transitions and
suggested that the stability of the period is best explained by assuming an underlying binary
system. In G9.62$+$0.20E three methanol lines at 6.7, 12.2 and 107~GHz showed flaring 
\cite[(van der Walt et al. 2009)]{vanderwalt09} and a colliding-wind
binary (CWB) scenario is found to explain periodicity through variations in the seed photon
flux and/or the pumping radiation field \cite[(van der Walt
2011)]{vanderwalt11}. Follow-up studies were required in order to provide more details
about the source; VLBI imaging revealed the maser
distribution \cite[(Goedhart et al. 2005)]{goedhart05} and multi-epoch
observations enabled the direct 
estimation of its distance of 5.2$\pm$0.6~kpc via trigonometric parallax 
(\cite[Sanna et al. 2009]{sanna09}). Recently, \cite[Szymczak et al.
(2011)]{szymczak11} discovered a similar case of variability in
G22.357$+$0.066 that can also be explained by changes in the background
free-free emission. A period of 179~days was derived from single dish
monitoring. The time delays seen between maser features can be combined with the VLBI imaging
to construct the 3D structure of the maser region. Another example is
G33.641$-$0.228 where the 6.7~GHz methanol bursts originate 
from a region of 70~AU \cite[(Fujisawa et al. 2012)]{fujisawa12}.
The authors interpret this as coming from an impulsive energy release like a stellar
flare. By monitoring many different objects we may also find more 
newly appearing masers as was the case with 6.7~GHz emission in IRAS~22506$+$5944 
\cite[(Wu et al. 2009)]{wu09}.

Variability was also detected for other maser transitions. 
In G353.273$+$0.641 intermittent 22~GHz maser flare activity appeared to be
accompanied by structural changes, likely indicating that the excitation is linked to an
episodic radio jet \cite[(Motogi et al. 2011b)]{motogi11b}. 
\cite[Lekht et al. (2012)]{lekht12} presented a catalog of 22~GHz H$_2$O spectra 
monitored over 30~years towards G34.3$+$0.15 (aka W44C). They detected a
long-term variability with an average period
of 14~years and two series of flares that are likely associated with
some cyclic activity of the protostar in the UC~H~{\small II}. 1720~MHz OH masers
towards W75N also showed flaring, possibly related to the very dense molecular material
that is excited and slowly accelerated by the outflow \cite[(Fish et al.
2011a)]{fish11a}. A surprising event was registered in IRAS18566$+$0408 by
\cite[Araya et al. (2010)]{araya10}: the 4.8~GHz H$_2$CO maser showed flaring
and a period correlated with the 6.7~GHz methanol outbursts. Both regions are
separated spatially, so both phenomena likely indicate variations in the 
infrared radiation field, maybe related to periodic accretion events. 
%Moreover, 
%a 6.035~GHz OH maser flare was found but delayed to H$_2$CO flare 
%(Al-Marzouk et al.\ {\it accepted by ApJ}.

\section{Masers in low and intermediate-mass protostars}

Because low and intermediate-mass stars are more common and evolve more
slowly, they should in principle be easier to study, at closer distances, 
maybe less embedded and in less confusing environments. It should therefore be possible
to study the star-formation process in more detail. 
However, masers are not so common in these kinds of objects and appear possibly in
different stages. 
\cite[Bae et al. (2011)]{bae11} found that only 9 and 6\% of a sample
of 180 intermediate-mass YSOs showed 22~GHz water and 44~GHz methanol maser
emission, respectively. Water is likely related to the inner parts of
outflows and can be highly variable, while methanol is possibly associated with the 
interfaces of the outflows with the ambient dense gas.  
The detection rates of both masers rapidly decreases as the central
(proto)stars evolve and the excitations of the two masers appear closely
related.  The most embedded (Class 0-like) intermediate-mass YSOs known to date are all
associated with water masers (e.g., \cite[S\'anchez-Monge et al.
2010]{sanchez10}; de Oliveira Alves, these proceedings). However, an intriguing
object, IRAS~00117$+$6412 was found with water masers in MM2, where
no outflow seems present \cite[(Palau et al. 2010)]{palau10}. In order to understand this
case, more observations with the best sensitivity and resolution are needed. 

The 44 and 36~GHz Class~I methanol masers rarely appear in lower-mass YSOs.
\cite[Kalenskii et al. (2010a)]{kalenskii10a} found only four 44~GHz and one 
36~GHz maser, while no emission was detected towards the remaining 39 outflows. 
They also noticed the masers have lower luminosity compared to those in HMSFRs. Imaging of L1157 
indicates that the 44~GHz maser may form in thin layers of turbulent 
post-shock gas or in collapsing clumps (\cite[Kalenskii et al.
2010b]{kalenskii10b}; Kalenskii et al.\ these proceedings). 

\section{HMSFRs in the Galactic Centre and beyond the Galaxy}
Recent searches towards Sgr~A revealed that the 36~GHz Class~I methanol masers
correlate with NH$_3$(3,3) density peaks and outline regions of cloud-cloud
collisions, maybe just before the onset of local massive star formation
\cite[(Sjouwerman et al. 2010a)]{sjouwerman10a}. The 44~GHz masers correlate with
the 36~GHz locations, but less with the OH masers at 1720~MHz \cite[(Pihlst\"om et al.
2011a)]{pihlstrom11a}, which are associated with the interaction between 
the supernova remnant Sgr~A East and the interstellar medium
\cite[(Pihlst\"om et al. 2011b)]{pihlstrom11b}. One particularly interesting 
group of 44~GHz masers was found that do not overlap with any 36~GHz emission.
These may signal the presence of a hotter and denser environment than the material swept up from 
the shock, maybe related to advanced star formation 
\cite[(Pihlst\"om et al. 2011a)]{pihlstrom11a}. 

{\it Kilomasers} are masers beyond our Galaxy, with luminosities comparable to 
the brightest galactic maser, which either amplify a background AGN or
originate from star-forming regions. These Galactic analog H$_2$O masers 
have become a great tool in studies of 
young super-star cluster formation with high angular resolution.
They are detected in a few nearby galaxies only (e.g., \cite[Castangia et al.
2008]{castangia08}).
\cite[Brogan et al. (2010)]{brogan10} found that water masers in the Antennae
Galaxies are associated with star-formation, as they show kinematic and spatial
agreement with massive and dense CO molecular clouds. The various early stages of
star formation in the components of the Antennae Galaxies was confirmed by \cite[Ueda et al. (2012)]{ueda12}. 

The Large and Small Magellanic Clouds were also searched for maser emission. 
\cite[Green et al. (2008)]{green08} detected four 6.7~GHz methanol (of which
one is new) and two 6.035 GHz~OH (of which one is new) masers in the LMC. Both
transitions indicate much more modest maser populations compared to the Milky Way,
likely originating from lower oxygen and carbon abundances. Moreover,
\cite[Ellingsen et al. (2010)]{ellingsen10} found the first 12.2~GHz methanol
maser towards the LMC. They also detected  22~GHz water and 6.7~GHz
methanol masers that are associated with more luminous and redder YSOs.
The first 6.7~GHz methanol spectrum towards Andromeda Galaxy (M31) was presented by 
\cite[Sjouwerman et al. (2010b)]{sjouwerman10b}.
More on kilo- and also mega-masers is presented by e.g.\ Tarchi (these proceedings).

%HH

\section{Magnetic field}
Masers are a particularly unique tool for studying the magnetic field studies in HMSFRs (Vlemmings, these proceedings). This is an important subject in star formation,
as the magnetic field could be a dominant force in the process by supporting the
molecular cloud against gravitational collapse, regulating the accretion
and shaping the outflows as has been argued for Cepheus~A (\cite[Vlemmings et al. 2006]{vlemmings06}). 
We have already mentioned the work by \cite[Green et al.(2007)]{green07} who
found a shock front in the form of a torus/ring around the YSO in ON~1. That
scenario is also supported by the measured polarization angles of the masers.
A magnetic field strength of a few mG was detected through the Zeeman splitting of OH and
methanol masers.
Linearly and circularly polarized emission of 22~GHz water masers was used for measuring the
orientation and strength of magnetic fields in W75N. The magnetic fields around the
young massive protostar VLA2 are found to be well ordered around an expanding gas shell 
\cite[(Surcis et al. 2011b)]{surcis11b}. 
In NGC~7538--IRS~1 the water masers did not show significant Zeeman
splitting while the 6.7~GHz methanol
masers indicated a possible range of magnetic field strength of 50~mG
$< | B_{\parallel} | <$ 500~mG,
depending on the value of Zeeman-splitting coefficient. These masers
likely are related to the outflow and the interface between infall and the
large-scale torus,
respectively \cite[(Surcis et al. 2011a)]{surcis11a}.

\section{Summary}
It is clear that many new results and discoveries have been obtained in our attempts
to understand star formation and the associated masers. Are we
closer to answer the key-questions? What is the current state of masers in SFRs?

We note, that:
\begin{itemize}
\item{The time sequence for masers seems to take shape and is
confirmed by (most) new methanol transition measurements, however, some issues still
exists,}
\item{Convergence can be seen on the issue where water, methanol (both Classes), OH and
SiO arise, but testing the hypotheses with high-resolution observations and at other
wavelengths is critical,}
\item{One may hope that in synergy with ALMA the role of masers to study
small scale dynamics will be strengthened,}
\item{Monitoring programs are starting to give important clues about co-evolving
binaries,}
\item{There should be more focus given to low-mass stars,}
\item{More work on models to get accurate physical conditions is needed.}
\end{itemize}

We started the review with the popular statement that {\it masers are an unique tool 
in star-formation studies} and in the end we are more convinced 
that they really are. In order to progress
the use of masers and solving the detailed questions
we {\it just} need better instruments and of course the patience to work on data to
obtain more and more interesting and even surprising results. We all can wait
(and work) with curiosity for the further discoveries that will be
presented in the next maser symposium.
\begin{acknowledgements}
AB acknowledges support by the Polish Ministry of Science and
Higher Education through grant N N203 386937. 
\end{acknowledgements}

%\begin{discussion}

%\discuss{XXX}{?}

%\discuss{van Langevelde}{}

%\end{discussion}

\end{document}